# Calculating Cognitive Augmentation – A Case Study

Ron Fulbright

University of South Carolina Upstate
Spartanburg, SC USA
rfulbright@uscupstate.edu

**Abstract.** We are entering an era in which humans will increasingly work in partnership and collaboration with artificially intelligent entities. For millennia, tools have augmented human physical and mental performance but in the coming era of cognitive systems, human cognitive performance will be augmented. We are only just now beginning to define the fundamental concepts and metrics to describe, characterize, and measure augmented and collaborative cognition. In this paper, the results of a cognitive augmentation experiment are discussed and we calculate the increase in *cognitive accuracy* and *cognitive precision*. In the case study, cognitively augmented problem solvers show an increase of 74% in cognitive accuracy—the ability to synthesize desired answers—and a 27% increase in cognitive precision—the ability to synthesize only desired answers. We offer a formal treatment of the case study results and propose cognitive accuracy and cognitive precision as standard metrics to describe and measure human cognitive augmentation.

## 1    Introduction

With recent advances in artificial intelligence (AI) and cognitive systems (cogs), we are at the beginning of a new era in human history in which humans will work in partnership with artificial entities capable of performing high-level cognition rivaling or surpassing human cognition. The new era will see human cognitive performance augmented by working with such artificial entities—*human cognitive augmentation*. Needed is a way to measure how cognitively augmented a human is by virtue of working with artificial entities. We propose two new metrics: *cognitive accuracy* and *cognitive precision*. Cognitive accuracy measures the ability to produce a desired result. Cognitive precision measures the ability to produce only the desired result and not any undesired results.

We can envision the situation as shown in Fig. 1. Together, a human and an artificial entity work together, forming a virtual entity, with the goal of performing a cognitive

task—the transformation of information in an input form to a desired output form. We can compare this situation to a human performing all cognition alone without the help of an artificial entity. The human involved in collaboration with the cog is cognitively augmented and therefore able to perform at a higher level than the human working alone. However, the question is how can we measure the degree of cognitive augmentation?

The result of any cognitive process can be either the desired result (or close to it) or an undesired result. We define *cognitive accuracy* ($C_A$) as the propensity to produce the desired result. We define *cognitive precision* ($C_P$) as the propensity to not produce something other than the desired result. Note, these are not necessarily equivalent to "correct" and "incorrect" results. Often, the result of cognitive processing cannot be labeled as correct or incorrect. For example, asking a person what things in life are important to them will generate a number of answers. It is not possible to determine if one of those answers is correct and the rest incorrect. However, we can identify a particular answer as being the one we desire. Once we have chosen the target, we can calculate accuracy and precision of any set of answers relative to the target.

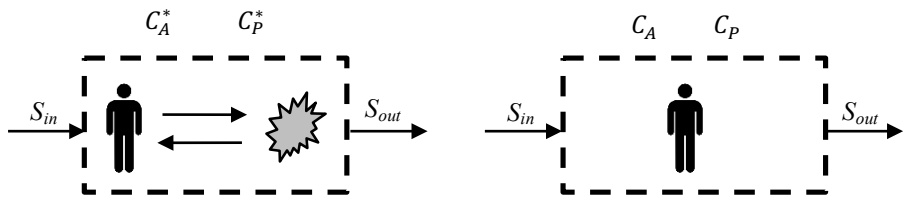

**Fig. 1.** In the coming era, human cognitive performance will be enhanced by partnering with artificially intelligent entities. The human/artificial ensemble will achieve a higher cognitive accuracy and cognitive precision than a human working alone.

This paper presents the results of a case study in which students were given an innovation problem and asked to synthesize as many solutions as they could think of in a period of time. Some students received expert advice in the form of suggested concepts pertaining to a class of preferred solutions. Other students received no assistance. Students receiving the expert advice represent cognitively augmented humans and students receiving no assistance represent non-augmented humans. Results show cognitively augmented students achieved a 74% increase in cognitive accuracy and a 27% increase in cognitive precision.

## 2    Previous Work

The idea of enhancing human cognitive ability with artificial systems is not new. In the 1640s, mathematician Blaise Pascal created a mechanical calculator called the Pascaline [1]. Thousands of years before this, mechanical devices such as the abacus aided basic arithmetic operations. Throughout history, humans have created thousands of such devices. Using these devices, a human is able to perform mathematical calculations difficult or impossible for an unaided human. These devices augment human mental performance, but the human still does all the thinking.

In the 1840s, Ada Lovelace was among the first to envision a machine performing a human task—musical composition [2, 3]. Lovelace imagined the machine composing the music, not a machine enhancing a human's ability to compose music. However, ideas like this were a century before their time. In the 1940s, Vannevar Bush envisioned a system called the Memex and discussed how employing associative linking could enhance a human's ability to store and retrieve information [4]. Similar to the above-mentioned calculating devices, the Memex made the human more efficient but did not actually do any of the thinking on its own.

In 1950, Turing discussed whether or not machines themselves could think and offered the "imitation game" as a way to decide if a machine is exhibiting intelligence [5]. Since coining the phrase artificial intelligence (AI) in 1955, several generations of researchers have sought to create an artificial system capable of human-like intelligence [6]. Also in the 1950s, Ross Ashby coined the term *intelligence amplification* maintaining human intelligence could be synthetically enhanced by increasing the human's ability to make appropriate selections on a persistent basis [7]. But here again, the idea is the human does all of the thinking. The synthetic aids just make the human more efficient.

In the early 1960s, Engelbart and Licklider envisioned human/computer symbiosis. Licklider imagined humans and computers becoming mutually interdependent, each complementing the other [9]. However, Licklider envisioned the artificial aids merely assisting with the preparation work leading up to the actual thinking which the human would do. Engelbart's H-LAM/T framework described the human as a part in a multi-component human/computer system allowing human and artificial systems to work together to perform problem-solving tasks [8]. Through the work of Engelbart's Augmentation Research Center, and other groups in the 1950s and 1960s, many of the devices we take for granted today were invented as "augmentation" tools including: the mouse, interactive graphical displays, keyboards, trackballs, WYSIWYG software, email, word processing, and the Internet. However, while making it easier for the human to think and perform, none actually do any of the thinking themselves. In essence, these are just modern versions of devices aiding human mental activity.

Recently, researchers have discussed entities capable of performing cognition on their own. One branch of AI has sought to develop semi-autonomous *intelligent software agents* to act on behalf of a user or other program [10, 11]. These agents are designed to interact as if they were human but also perform on their own without supervision from the human user. The concept of the agent—a self-contained, interactive and concurrently executing object, possessing internal state and communication capability—can be traced to the Hewitt's Actor model [12]. Software agents act autonomously and only occasionally communicate with the human user. However, the field of *human-autonomy teaming* has studied real-time interaction between humans and artificial systems. One active area involves military applications. Constraints must be met though because combat requires systems to respond rapidly and efficiently while attaining mission objectives [13]. One goal of this research is an Autonomous Squad

Member where a human squad member, either in a military or law enforcement setting, is assisted by an autonomous agent in mission environments [14]. NASA has researched the idea of having fewer human operators on long space flights by using artificial intelligence [15, 16].

However, these are all highly-specialized applications. What about the average person? Thirty years ago, Apple, Inc. envisioned an intelligent assistant called the Knowledge Navigator [17]. The Knowledge Navigator was an artificial executive assistant capable of natural language understanding, independent knowledge gathering and processing, and high-level reasoning and task execution. The Knowledge Navigator was envisioned as a colleague anyone could work with. The Knowledge Navigator concept was well ahead of its time, however, some of its features are seen in today's voice-controlled "digital assistants" such as Siri, Cortana, and Amazon Echo.

More recently, [Forbus, 18] described companion cognitive systems as software collaborators helping their users work through complex arguments, automatically retrieving relevant precedents, providing cautions and counter-indications as well as supporting evidence. Companions assimilate new information, generate and maintain scenarios and predictions, and continually adapt and learn about the domains they are working in, their users, and themselves. Companions, operate in a limited domain however achieve expert-level performance in that domain, often exceeding that of a human expert.

[Langley, 19] challenged the cognitive systems research community to develop a synthetic entertainer, a synthetic attorney, and a synthetic politician as a way to drive future research in integrated cognitive systems. The vision here is to develop a virtual human. We maintain the goal should be not to create a virtual human capable of being an entertainer, an attorney, or a politician, but rather create a cognitive system capable of expert-level performance in entertainment, a different cognitive system capable of exhibiting expert performance in a subfield of law, and a cognitive system capable of expert politicking.

[Fulbright, 41-45] foresees the creation a cognitive system for virtually any human endeavor and maintains the democratization of expertise will change the way we live, work, and play over the next several decades much like the computer and Internet have changed our lives over the last few decades.

A significant step toward this vision occurred in 2011, when a cognitive computing system built by IBM, called Watson, defeated two of the most successful human *Jeopardy!* champions of all time [20]. Watson received clues in written natural language and gave answers in natural spoken language. Watson's answers were the result of searching and deeply reasoning about millions of pieces of information and aggregation of partial results with confidence ratios. Watson was not programmed to play *Jeopardy!* Instead, Watson was programmed to *learn* how to play *Jeopardy!* which it did in many training games with live human players before the match [21, 22]. Watson *practiced* and achieved expert-level performance within the narrow domain of playing *Jeopardy!* Watson represents a new kind of computer system called a cognitive system [23, 3].

Instead of replacing humans, cognitive systems seek to act as partners with and alongside humans. John Kelly, Senior Vice President and Director of Research at IBM describes the coming revolution in cognitive augmentation as follows [24]:

*The goal isn't to… replace human thinking with machine thinking. Rather…humans and machines will collaborate to produce better results – each bringing their own superior skills to the partnership. The machines will be more rational and analytic – and, of course, possess encyclopedic memories and tremendous computational abilities. People will provide judgment, intuition, empathy, a moral compass and human creativity."*

Cognitive systems, whether or not they are human or artificial, process and transform information. To measure and characterize cognition then, we must consider how to measure information and quantify information processing. Most information content metrics devised so far key off of the structure of the information being processed. In 1948, Claude Shannon developed the basis for what has become known as *information theory* [25-27]. Shannon, like Hartley before him, equates order/disorder and information content. In the 1960's, Ray Solomonoff, Gregory Chaitin, Andrey Kolomogorov and others developed the concept of *algorithmic information theory* (Kolmogorov-Chaitin complexity) as a measure of information [28-31]. The algorithmic information content, $I$, of a string of symbols, $w$, is defined as the size of the minimal program running on the universal Turing machine generating the string. This measure of information concerns the complexity of a data structure as measured by the amount of effort required to produce it. A string with regular patterns can be "compressed" and produced with fewer steps than a string of random symbols which requires a verbatim listing symbol by symbol. Like the entropic measures described above, this description equates order/disorder to information content, although in a different manner by focusing on the computational resources required. In 1990, Tom Stonier suggested an exponential relationship between entropy, $S$, and information [32-34]. Stonier also maintained information content is dependent on the *structure* present and uses Shannon's entropy to provide the measure of that structure.

Renaldo Vigo has recently proposed a new kind of information theory, *generalized representational information theory* (GRIT) [35, 36]. Key to GRIT is how humans learn concepts from information. Empirical evidence shows human concept extraction is based on the detection of patterns—invariance in the information. Vigo's *generalized invariance structure theory* (GIST) maintains it is easier to extract a concept from information with less variance (more similarity between elements) than it is from information with a more variance (less similarity between elements). A key measure of information is then the *structural complexity*—a different measure of structure than Shannon entropy. Values calculated with GIST formulae agree with empirical evidence from human trials.

Fulbright has proposed several metrics based on GRIT and GIST for describing cognitive processing [37-39]. Cognitive work, is an accounting of all changes in structural complexity caused by a transformation of information. Cognitive work is a measure of the total effort expended in the execution of a cognitive process. When humans (H) and artificial entities, or cogs, (C) work together, each are responsible for some amount of change (cognitive gain (G)) and each expend a certain amount of cognitive work (W):

$$W^* = W_H + W_C \qquad G^* = G_H + G_C \qquad (1)$$

Given that we can calculate the individual cognitive contributions, it is natural to compare their efforts. In fact, doing so yields the *augmentation factor, $A^+$*:

$$A_W^+ = \frac{W_C}{W_H} \qquad A_G^+ = \frac{G_C}{G_H} \qquad (2)$$

Note humans working alone without the aid of artificial entities are not augmented at all and have an $A^+ = 0$. If humans are performing more cognitive work than artificial entities, $A^+ < 1$. This is the world in which we have been living so far. However, when cogs start performing more cognitive work than humans, $A^+ > 1$ with no upward bound. That is the coming cognitive systems era. Fulbright defines other efficiency metrics by comparing cognitive gain and cognitive work to each other and to other parameters such as time, $t$, and energy, $E$.

However, all of these metrics focus on the microscopic features of the information itself—structure. As such, they are based on quantities difficult, if not possible, to measure and calculate. This renders these metrics interesting conceptually but useless in a practical sense. In this paper, we discuss two macroscopic metrics: *cognitive accuracy* and *cognitive precision*. Instead of focusing on characteristics of the information, these two metrics focus on the results of the cognitive processing making them easy to calculate and useful across multiple domains.

## 3   Cognitive Accuracy and Cognitive Precision

To define the notions of *cognitive accuracy* and *cognitive precision*, we first model the students in the class with a general information machine (GIM) [40]. A GIM is a stochastic Turing machine accepting information as an input and producing information as an output. In a traditional Turing machine, rules specify symbol transitions dictating a deterministic transformation of an input to a specific output. However, the rules in a GIM are stochastic in nature. Each transition rule is associated with a probability rather than being a deterministic certainty. Therefore, for a given input, the GIM's output may vary with each run. Over a number of runs, given the same input, a set of outputs ($C$) is created. The pattern of outputs in $C$ is determined by the randomness of the probabilities within the GIM and denoted as $\lambda$ as shown in Fig. 2.

If the GIM is truly random, the outputs in $C$ are evenly distributed with the average probability of each output, $c$, being $1/|C|$ where $|C|$ is the cardinality of $C$ or simply the number of different outputs. If the GIM is truly deterministic (such as that of a traditional Turing machine), one and only one output will be generated 100% of the time. Of course, the probability of that output is 100% and the probability of any other possible output is zero. If, however, the randomness of the GIM is an intermediate value a pattern of outputs will emerge. Some of these outputs will be very similar to other outputs and can be grouped together into subsets of $C$. The probability of a subset can be calculated by comparing the cardinality of the subset with the cardinality of $C$.

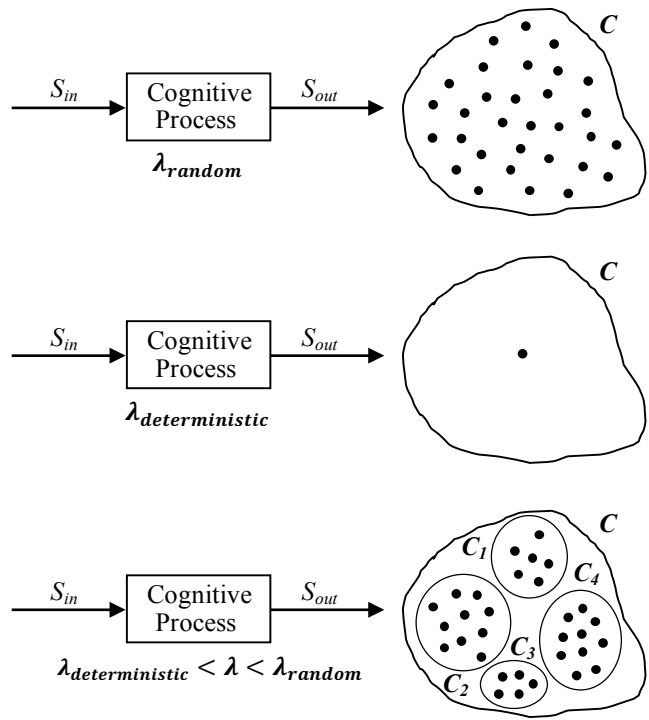

**Fig. 2.** Cognitive processing modeled as stochastic manipulation of information.

Critical is the distribution pattern of ***C***. If we choose one of the subsets in ***C*** as the preferred type of output we can characterize any distribution pattern based on the ideas of *accuracy* and *precision* as shown in Fig. 3. Accuracy involves the propensity to hit the preferred subset. Precision involves the propensity to hit *only* the preferred subset. The goal, of course, is for every output to fall within the preferred subset (upper right quadrant). This represents high accuracy and high precision. It is possible for outputs to be very similar to each other (forming a tight cluster) but not falling within the preferred subset (lower right quadrant). This represents high precision but low accuracy. Outputs centered on the preferred subset but not tightly clustered (upper left quadrant) represents high accuracy but low precision. Outputs with low accuracy and low precision (lower left quadrant) have only accidental relationship to the preferred subset.

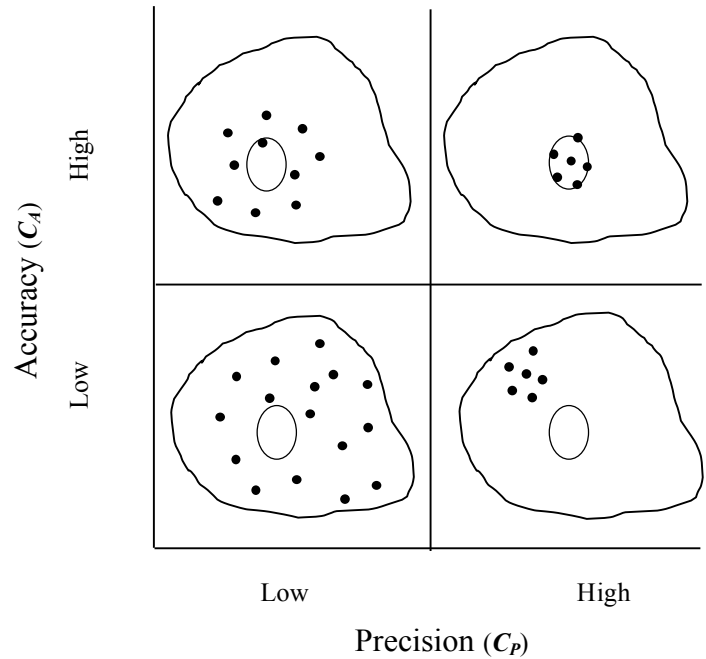

**Fig. 3.** Precision and accuracy relative to a target subset.

Since the outputs in the model are the result of cognitive processing, we call these two measures *cognitive accuracy ($C_A$)* and *cognitive precision ($C_P$)*.

## 4   The Case Study

For the case study, an innovation problem was given to a classroom of students registered into the INFO 307: Systematic Innovation course at the University of South Carolina Upstate. INFO 307 teaches students an innovation methodology called I-TRIZ. This test was done on the first day of two consecutive semesters before any instruction took place. This was done so as to not bias any results with the effect of learning the methodology. One semester consisted of 25 students and the second semester consisted of 21 students. Students were given ten minutes to write down as many solutions as they could think of to the following problem:

*Problem: Skeet shooting is a recreational and competitive activity where participants, using shotguns, attempt to break clay targets mechanically flung into the air from fixed stations at high speed from a variety of angles. The problem is the shattered skeet litter the grass field harming the grass. As you know, grass needs sun, water, and nutrients to be healthy. The skeet fragments prevent water and sun from reaching the grass and diminish the healthy nutrients in the soil after they eventually dissolve.*

The I-TRIZ methodology employs a collection of innovative concepts, called *operators*, gleaned from the study of millions of patents. The operators represent a distillation of human innovative thought and therefore represent a collection of expert knowledge. In the I-TRIZ methodology, operators are used to stimulate thinking and inspire solutions to problems. The case study was designed to demonstrate the effect operators have on problem solving ability even for those without training in the methodology. One-third of the students were given the above problem statement without any other information at all (no operators). One-third of the students were given the problem statement and a list of five operators (OPS 1) and one-third of the students were given the problem statement and a list of five additional operators (OPS 2) for a total of ten operators. The operators given were:

OPS 1 = 1. exclude the source
2. use a disposable object
3. apply liquid support/Introduce a liquid
4. inversion (apply the opposite)
5. phase transformation (freeze/melt/boil; solid/liquid/gas/plasma)

OPS 2 = 1. transform a substance to a fluid state
2. self-healing (system corrects itself)
3. formation of mixtures
4. transform the aggregate state to eliminate a harmful effect
5. resources - modified water

These operators were chosen for a reason. Based on previous experience, solutions to this problem generally fall into three distinct categories:

*F*: modifying the field/cleaning up the field
*T*: modifying the target (skeet)
*G*: modifying the gun or bullet.

Solutions in the field category (*F*) include: covering the field with a tarp or net, various ways of cleaning up the field, and relocating to a location without a grass field.

Solutions in the target category (**T**) include: using targets made of biodegradable material, using targets made of fertilizer and other nutrients good for the grass, and enhancing the rapid dissolvability of the targets. Solutions in the gun category (**G**) include: using a different kind of gun or bullet, and using a laser or electromagnetic gun/target.

For the case study, the modifying the target (**T**) category of solutions was chosen as the preferred type of solution because solutions in the other two categories are considered obvious solutions (the first ones most people think of off the top of their head). Operators in OPS1 and OPS2 shown above were chosen with the intent of driving student thinking toward solutions in the **T** category.

The results of the case study are shown in Fig. 4. A total of 128 solutions were created by the students. Of course, there were many duplicative solutions. In all, 11 different solutions were synthesized by the class over the two semesters. Each of the 128 solutions turned in by the students were classified into one of the three solution categories and the simple percentage for each category was calculated.

Students receiving only the problem statement and no operators produced solutions in the preferred subset just over one quarter of the time for an accuracy of, $C_A = 27\%$. Students receiving five operators (OPS 1) produced solutions in the preferred subset with an accuracy of $C_A = 40\%$. Students receiving ten operators (OPS 1 + OPS 2) achieved an accuracy of $C_A = 47\%$. Therefore, the operators increased the students' cognitive accuracy by (40%-27%)/27% = 48% (for OPS1) and (47%-27%)/27% = 74% (for OPS 1 + OPS 2).

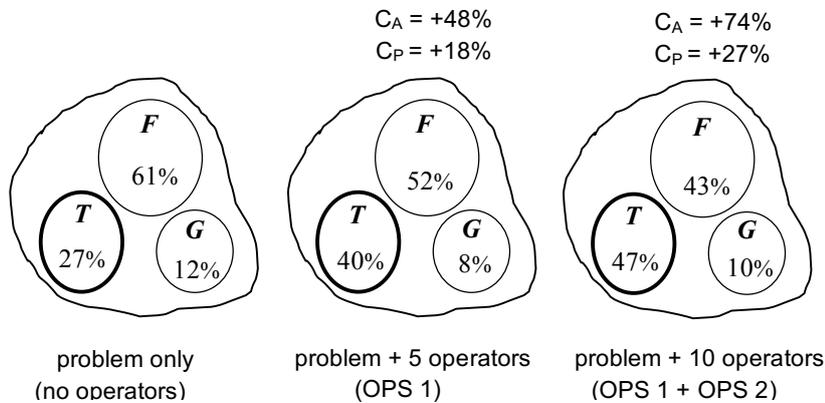

**Fig. 4.** The cognitive augmentation effect of operators on problem solving is demonstrated relative to the preferred type of solution **T**. Cognitive accuracy ($C_A$) was increased by 48% and 74% respectively. Cognitive precision ($C_P$) was increased by 18% and 27% respectively.

Non-augmented students tended to focus on the modifying the field (**F**) type of solution (61% of the solutions). Together with the modifying the gun/bullet (**G**) type of solution (12%), non-augmented students therefore missed the preferred subset 73% of the time. However, students augmented with the operators created far fewer solutions outside of the preferred subset: 52% + 8% = 60% (for OPS 1) and 43% + 10% = 53% (for OPS 1 + OPS 2). Therefore, augmented students achieved a reduction of the miss-rate (60% -73%)/73% = -18% (for OPS 1) and (53%-73%)/73% = -27% (for OPS 1 + OPS 2) representing an *increase in cognitive precision*.

Students were cognitively augmented by the operators. Even without formal training in the IPS methodology, students were able to create solutions of the preferred (non-obvious) type far in excess of students solving the problem without operators.

## 5    Conclusion

In the coming era of cognitive systems, cognition will be the result of human collaboration with artificially intelligent entities (cogs). Cogs will think on their own and offer expert advice to the human. The human will consider the expert advice and use it to enhance the quality of his or her own thinking. Together, the biological/artificial collaboration will yield a cognitive product greater than each of the entities could have produced on their own. Humans will be *cognitively augmented.* We have proposed two new metrics to describe the result of such cognitive augmentation. *Cognitive accuracy* is the ability of the human/artificial ensemble to create the desired output. *Cognitive precision* is the ability of the ensemble to create nothing but the desired output.

The results of a case study were presented. In the case study, student problem solving was augmented by considering problem solution concepts called operators. The cognitive accuracy of the cognitively augmented students was increased by as much as 74% meaning they were significantly more likely to create solutions of the preferred type. Furthermore, the cognitive precision of the group was increased by 27% meaning the students were less likely to create solutions not of the preferred type.

We believe the metrics of cognitive accuracy and cognitive precision are two ideas applicable to any study of cognition, biological or artificial. Since these ideas rely only on the results of cognitive processing and do not rely on the structure of the information involved nor on any particulars of how the cognition is performed these metrics can be used to compare and contrast results across multiple domains.

# References

1. Chapman, S. (1942). Blaise Pascal (1623-1662) Tercentenary of the calculating machine. *Nature,* London 150: 508–509.
2. Hooper, R. (2012). Ada Lovelace: My brain is more than merely mortal, New Scientist, Internet page located at https://www.newscientist.com/article/dn22385-ada-lovelace-my-brain-is-more-than-merely-mortal last accessed November 2015.
3. Isaacson, W. (2014). The Innovators: How a Group of Hackers, Geniuses, and Geeks Created the Digital Revolution, Simon & Schuster, New York, NY.
4. Bush, V. (1945). As We May Think, The Atlantic, July.
5. Turing, A. (1950). Computing Machinery and Intelligence, *Mind*, Vol. LIX, Issue 236, October 1950.
6. McCarthy, J., Minsky, M., Rochester, N., Shannon, C. (1955). A Proposal for the Dartmouth Summer Research Project on Artificial Intelligence. Stanford University Internet page locate at http://www-formal.stanford.edu/jmc/history/dartmouth/dartmouth.html last accessed November 2015.
7. Ashby, W.R. (1956). An Introduction to Cybernetics, Chapman and Hall, London.
8. Engelbart, D.C. (1962). *Augmenting Human Intellect: A Conceptual Framework, Summary Report* AFOSR-3233, Stanford Research Institute, Menlo Park, CA, October 1962.
9. Licklider, J.C.R. (1960). Man-Computer Symbiosis, *IRE Transactions on Human Factors in Electronics*, Vol. HFE-1, March.
10. Nwana, H.(1996) Software Agents: An Overview, retrieved October 9, 2018 from https://teaching.shu.ac.uk/aces/rh1/elearning/multiagents/introduction/nwana.pdf.
11. Schermer, B. W. (2007) Software Agents, Surveillance, and the Right to Privacy: A Legislative Framework for     Agent-Enabled Surveillance, Leiden, retreived October 9,2018 from https://openaccess.leidenuniv.nl/bitstream/handle/1887/11951/Thesis.pdf.
12. Hewitt, C., Bishop, P., Steiger, R. (1973), A Universal Modular Actor Formalism for Artificial Intelligence, WorryDream, retrieved October 9, 2018, from http://worrydream.com/refs/Hewitt-ActorModel.pdf
13. Barnes, M., Chen, J. Y. C., Hill, S., (2017), Humans and Autonomy: Implications of Shared Decision-Making for Military Operations, Army Research Laboratory, retrieved October 6, 2018, from http://www.arl.army.mil/arlreports/2017/ARLTR-7919.pdf
14. Chen, J. Y. C., Stowers, K., Barnes, M. J., Selkowitz, A. R., Lakhamni, S. G., (2017) Human-Autonomy Teaming and Agent Transparency, *HRI '17 Proceedings of the Companion of the 2017 ACM/IEEE International Conference on Human-Robot Interaction*, Association for Computing Machinery, March 2017.
15. Shivley, R. J.,Bandt, S. L. Lachter, J., Matessa, M., Sadller, G., and Battise, V.(2016). Application of Human Autonomy Teaming (HAT) Patterns to Reduce Crew Operations (RCO), National Aeronautics and Space Admisitration, retrieved September 17, 2018 from https://ntrs.nasa.gov/archive/nasa/casi.ntrs.nasa.gov/20160006634.pdf
16. Shivley, R. J.,Bandt, S. L. Lachter, J., Matessa, M., Sadller, G., Battise, V, and Johnson, W., (2018). Why Human Autonomy Teaming, ResearchGate, retrieved October 9,2018, from https://www.researchgate.net/publication/318182279_Why_Human-Autonomy_Teaming
17. Apple (1987). Knowledge Navigator, YouTube, retrieved April 2016 from https://www.youtube.com/watch?v=JIE8xk6Rl1w

18. Forbus, K. and Hinrichs, T. (2006). Companion Cognitive Systems: A Step Toward Human-Level AI, AI Magazine, Vol. 27, No. 2.
19. Langley, P. (2013). Three Challenges for Research On Integrated Cognitive Systems, Proceedings of the Second Annual Conference on Advances in Cognitive Systems,
20. Jackson, J., (2011). IBM Watson Vanquishes Human Jeopardy Foes, *PC World*. Internet page http://www.pcworld.com/article/219893/ibm_watson_vanquishes_human_jeopardy_foes.html last accessed May 2015.
21. Ferrucci, D.A. (2012). Introduction to "This is Watson," IBM J. Res. & Dev. Vol. 56 No. 3/4.
22. Ferrucci, D., Brown, E., Chu-Carroll, J., Fan, J., Gondek, D., Kalyanpur, A., Lally, A., Murdock, J.W., Nyberg, E., Prager, J., Schlaefer, N., and Welty, C. (2010). Building Watson: An Overview of the DeepQA Project, *AI Magazine*, Vol. 31, No. 3.
23. Wladawsky-Berger, I. (2013). The Era of Augmented Cognition, *The Wall Street Journal: CIO Report*, Internet page located at http://blogs.wsj.com/cio/2013/06/28/the-era-of-augmented-cognition/ last accessed May 2015.
24. Kelly, J.E. and Hamm, S. (2013). Smart Machines: IBMs Watson and the Era of Cognitive Computing, Columbia Business School Publishing, Columbia University Press, New York, NY.
25. Hartley, R.V.L. (1928). Transmission of information, *Bell System Technical Journal*, 7 (1928):535-563.
26. Shannon, C.E. (1948). A Mathematical Theory of Communication. *Bell System Technical Journal*. 27 (3): 379–423.
27. Weaver, W. and Shannon, C.E. (1949). *The Mathematical Theory of Communication*, University of Illinois Press, Urbana, Illinois 1949.
28. Chaitin, G. J. (1966). On the Length of Programs for Computing Finite Binary Sequences, *Journal of the Association of Computing Machinery* 13, 1966: 547-569.
29. Chaitin, G. J. (1977). Algorithmic Information Theory, *IBM Journal of Research and Development*, 21, 1977: 350-9,496.
30. Kolmogorov, A. N. (1965). Three Approaches to the Quantitative Definition of Information, *Problems of Information Transmission*, 1, 1965: 1-17.
31. Solomonoff, R.J. (1964). A Formal theory of Inductive Inference, *Information and Control* 7, 1964: 1-22, 224-54.
32. Stonier, T. (1990). *Information and the Internal Structure of Universe*, Springer-Verlag, London, 1990.
33. Stonier, T. (1992). *Beyond Information: The Natural History of Intelligence*, Springer-Verlag, London, 1992.
34. Stonier, T. (1997). *Information and Meaning: An Evolutionary Perspective*, Springer, Berlin, 1997.
35. Vigo, R. (2013). Complexity over Uncertainty in Generalized Representational Information Theory (GRIT), *Information*, Vol. 4.
36. Vigo, R. (2015). *Mathematical Principles of Human Conceptual Behavior*, Psychology Press, NY, New York, ISBN: 978-0-415-71436-5.
37. Fulbright, R. (2016). How Personal Cognitive Augmentation Will Lead To The Democratization of Expertise, Fourth Annual Conference on Advances in Cognitive Systems, Evanston, IL, June 2016. Available at http://www.cogsys.org/posters/2016, last retrieved January 2017.
38. Fulbright R. (2017). Cognitive Augmentation Metrics Using Representational Information Theory. In: Schmorrow D., Fidopiastis C. (eds) *Augmented Cognition. Enhancing Cognition and Behavior in Complex Human Environments*. AC 2017. Lecture Notes in Computer Science, vol 10285. Proceedings of HCI International 2017 conference Springer.

39. Fulbright R. (2018). On Measuring Cognition and Cognitive Augmentation. In: Yamamoto, S. and Mori, H. (eds) *Human Interface and the Management of Information*. LNCS 10904. Proceedings of HCI International 2018 conference. Springer.
40. Fulbright, R. (2002). Information Domain Modeling of Emergent Systems, Technical Report CSCE 2002-014, May 2002, Department of Computer Science and Engineering, University of South Carolina, Columbia, SC.
41. Ron Fulbright, "On Measuring Cognition and Cognitive Augmentation," HCI International 2018, Las Vegas, NV, July 2018.
42. Ron Fulbright, "Cognitive Augmentation Metrics Using Representational Information Theory," HCI International 2017, Vancouver, July 2017.
43. Ron Fulbright, "ASCUE 2067: How We Will Attend Posthumously," Proceedings of the 2017 Association of Small Computer Users in Education (ASCUE) Conference, June 2017.
44. Ron Fulbright, "How Personal Cognitive Augmentation Will Lead to the Democratization of Expertise," The Fourth Annual Conference on Advances in Cognitive Systems, June 2016.
45. Ron Fulbright, "The Cogs Are Coming: The Coming Revolution of Cognitive Computing," Proceedings of the 2016 Association of Small Computer Users in Education (ASCUE) Conference, June 2016.